# Science, Religion and the Teaching of Evolution

Pierre C. Hohenberg,
New York University
August 2010

Abstract

This essay discusses the relationship between science and religion, specifically the controversy elicited by an article by the philosopher Thomas Nagel criticizing the scientific establishment for "ruling out intelligent design as beyond discussion". He also criticizes the judge's decision in *Kitzmiller vs. Dover* ruling out discussion of intelligent design in science classrooms in public schools. A defense of the thesis of Stephen Gould that science and religion represent 'nonoverlapping magesteria' (NOMA) is presented.

I.   Introduction

A prevalent but by no means universally held view of the relationship between science and religion, exemplified by Stephen Gould's "nonoverlapping magisteria" (NOMA) principle (Gould, 1997), is that they are fully compatible and thus can peacefully coexist. This view is exemplified by a National Academy of Sciences (2008) pamphlet on Science Evolution and Creationism and a related book by one of its leading authors (Ayala, 2007) and it is also eloquently expressed by a pronouncement of Pope John Paul II (1996) which lends religious authority to the NOMA principle. Nevertheless, apart from the proponents of creationism or intelligent design, who should not be taken seriously (see below), there are dissenting voices on both the science and the non-science sides. In the former category I would cite Richard Dawkins (2006), who considers NOMA to be at best a misguided form of appeasement. On the other side I will cite Thomas Nagel (2006, 2008) who has criticized Dawkins's views on religion and, more to the point, has also criticized the Dover decision on the teaching of evolution in the schools as implying that evolutionary theory cannot be taught in an intellectually responsible way.

In the present essay I argue that what is missing in the above 'prevalent' view is a proper *justification* for the difference between the magisteria of science and religion.

It is precisely this justification which in my view answers the objections of Dawkins and Nagel. Briefly, the difference arises from the essence of Science as a collective human endeavor of assembling *public knowledge*, universally applicable but forever subject to change via improvement and refinement (Hohenberg, 2010). Religion, on the other hand, involves private or at least local knowledge.

II.   Science, science and non-Science

As explained in more detail elsewhere (Hohenberg 2010), it is useful to distinguish between "(lower-case) science" as a human pursuit carried out by scientists, with all of its contingencies , controversies and contradictions, and the public knowledge that emerges from this activity, which we refer to as "(upper-case) Science". Newton and Einstein, not to speak of the present author, produced science. Their work (especially the first two) eventually emerged into Science, at which point it no longer "belonged" to them, any more than *2 + 2 = 4* belongs to the originators of arithmetic.

In addition to science and Science there is "non-Science", which encompasses all the rest of human knowledge, i.e. art, literature, politics, ethics, religion, technology, etc. All of these areas, including (lower case) science, are in one way or the other associated with people, either as individuals or in groups, who must "own", "believe" or "commit to" the relevant knowledge. Religion, in particular cannot exist without believers, but our thesis is that in some sense Science can.

Questions that are amenable to universal answers are the proper domain of Science, where the answers coexist with a certain amount of ignorance, i.e. they represent partial and often provisional knowledge which has nevertheless attained a level of general acceptance and can thus be termed "objective". Questions that have little or no expectation of leading to universal answers are the domain of non-Science, which encompasses some of the most urgent and interesting questions faced by mankind. Examples are:

- How do I find meaning in my life?
- Is this good or bad?
- How do I relate to my fellow human beings?
- What is the best political system?
- Does God exist?

Providing answers to these questions involves an act of individual or collective will, which gives the answers a subjective character. The above list emphasizes moral or psychological issues but such pursuits as medicine or technology are also included in non-Science, even though they rely in considerable measure on scientific knowledge. A doctor's job is to cure patients and that involves an interaction between individuals that clearly transcends science. The same can be said of teaching physics or teaching evolutionary biology!

The significance of nonscientific knowledge is brought home by the observation that scientists do not seem to be more moral or happier than their nonscientist colleagues (e.g. literature professors or orchestra musicians), even though scientists possess so much more "reliable" and universal knowledge.

III.    Science and Religion

The prevalent view we described in the Introduction, with which we agree, states that Science and religion are responses to different classes of questions: those that are expected to have universal (i.e. anonymous) answers belong to Science and they remain a mix of (objective) knowledge and ignorance, in varying proportions; those questions whose answers involve belief, commitment or some act of will belong to non-Science and an important subclass are questions of a religious, or existential nature. From this point of view every human being is "religious", in that he or she confronts such questions either explicitly or implicitly at some point, even an atheist. The important point is that the answers must be "owned" by a "believer", be it an individual or a group. Atheism, humanism or agnosticism are religious answers from this point of view.

The notion of a universal religion, on the other hand, seems curiously out of date in the $21^{st}$ century. It is embraced by the type of people we wish to lock up (e.g. terrorists) or at least severely curtail (e.g. religious fundamentalists). In some sense the claims to religious universality have been replaced in the past 400 years, by claims to universality for Scientific knowledge, albeit at the price of relinquishing the ambition to answer many questions that people really care about.

The difference between Science and religion is well illustrated by the concept of *ecumenism*, an entirely respectable principle which most religious leaders profess to accept, even if at times reluctantly. The notion of scientific ecumenism, on the other hand, i.e. finding a mode of peaceful coexistence between mutually contradictory theories or results, is an absurdity. Science seeks to eliminate, or at least resolve contradictions and considers remaining ones as part of the "sea of ignorance" in which Scientific knowledge necessarily bathes (Hohenberg, 2010).

There is a famous saying, of Richard Dawkins I believe, though I have not found the reference, that the difference between him and a fundamentalist Christian, say, is negligible since they both reject the thousands of gods proposed by mankind, and he (Dawkins) merely goes "one god further". The joke is based on the mistaken or at least obsolete notion of universal religion. In fact, however, all 1001 gods can coexist as part of the diversity of human experience and Dawkins's rejected "last god" is just an expression of his religion, atheism.

How then can (or should) Science coexist with religion? The simplest solution is for religion to ignore Science as irrelevant to the concerns of most believers, a solution that is in practice adopted by a majority of the world's population. It is not, however, acceptable to educated persons who aspire to share in the universal knowledge claimed by Science. For those individuals religion should seek to occupy the vast areas that are unattainable by Science rather than invade the territory of Science. In a sense this is a "god of the gaps", but the gaps are not the ignorance of "not yet Science". Rather they are the gaps of "never Science" because the answer requires an act of will or personal commitment. Finding the proper balance between religious and Scientific truths is certainly a challenge for religious teaching and practice, but there are many examples of success, as illustrated for example by the statements in the National Academy of Science (2008) pamphlet, or in the writings of Francisco Ayala (2007) mentioned earlier, as well as the writings of Freeman Dyson (e.g. Dyson, 2004), all consistent with the NOMA principle.

IV.     Reductionism, Materialism, Naturalism and Supernaturalism

The rejection of NOMA by both Dawkins and Nagel is framed in terms of a *philosophical* debate concerning rational arguments for so-called "natural religion". Dawkins goes even further, treating the question of the existence of god as a *scientific* issue (Dawkins, 2006, p. 48):

> " Either he (god) exists or he doesn't. It is a scientific
> question; one day we may   know the answer and meanwhile
> we can say something pretty strong about the probability."

Nagel (2006) is willing to engage in this debate when he juxtaposes "Dawkins's physicalist naturalism… (to) the god hypothesis...", or remarks that "the conflict between religious belief and atheism takes the form of a scientific disagreement." Note, however, that he does remark elsewhere that "…this whole dialectic leaves out another possibility, namely that there are teleological principles that are explained neither by intentional design nor by purposeless physical causation, principles that therefore provide an independent end point of explanation for the existence and form of living things." We shall comment on this alternative below.

It is our view that this philosophical debate, interesting as it may be, has little to do with either Science or religion as we have characterized them. Scientific debates are resolved by the (slow) process of emergence alluded to above, in which belief plays little role. Religious issues, on the other hand, are not primarily resolved by rational arguments. From our point of view, philosophical debates are neither like Science nor religion, but more akin to (lower case) science (in that they are based on rational arguments), with the important difference that their end point is not likely to be emergence into objective, i.e. universal, knowledge. In practice the issues are sufficiently complex and multi-faceted that opposing points of view are expected to persist for a long time if not for ever. This is not to say that the philosophy of science cannot illuminate scientific debates and influence the Science of the future as they have the Science of the past, but rather that a direct and universal resolution of the questions under debate is not expected.

The philosophical issue that is relevant here concerns the proper role of physicalist reductionism, materialism, naturalism and supernaturalism in one's world view and in the basic paradigm of Science. We believe that it is stated and unstated assumptions about those issues that are at the heart of the controversies surrounding science, religion and evolution. Let us begin by defining more precisely the different terms and points of view.

*Physicalist reductionism* (also referred to as "grand reductionism") is well expressed by Weinberg (1995):

> "Grand reductionism is … the view that all of nature is the way it is (with certain qualifications about initial conditions and historical accidents) because of simple universal laws, to which all other scientific laws may in some sense be reduced… One can illustrate the reductionist world view by imagining all the principles of science as being dots on a huge chart, with arrows flowing into each principle from all the other principles by which it is explained… We say that one concept is at a higher level or a deeper level than another if it is governed by principles that are further from or closer to this common source… The reductionist program of physics is the search for the common source of all explanations… But phenomena like mind and life do emerge. The rules they obey are not independent truths, but follow from scientific principles at a deeper level,… The nervous systems of George and his friends have evolved to what they are entirely because of the principles of macroscopic physics and chemistry, which in turn are what they are entirely because of principles of the standard model of elementary particles… But there are no principles of chemistry that simply stand on their own, without needing to be explained reductively from the properties of electrons and atomic nuclei, and in the same way there are no principles of psychology that are free-standing, in the sense that they do not need ultimately to be understood through the study of the human brain, which in turn must ultimately be understood on the basis on physics and chemistry."

A similar description is given by Dawkins (1986):

" The hierarchical reductionist… explains a complex entity at any particular level in the hierarchy of organization in terms of entities only one level down in the hierarchy, entities which, themselves, are likely to be complex enough to need further reducing to their own component parts; and so on."

The well-known biologist and naturalist E. O. Wilson has formulated an ambitious program of unification of all aspects of human knowledge (Wilson, 1998, p. 307):

" The gaps of greatest potential include the final unification of physics, the reconstruction of living cells, the assembly of ecosystems, the coevolution of genes and culture, the physical basis of mind, and the deep origins of ethics and religion."

He stresses, however, that this unification is to be based on grand reductionism (Wilson, 1998, p. 305):

" The central idea of the consilience world view is that all tangible phenomena, from the birth of stars to the workings of social institutions, are based on material processes that are ultimately reducible, however long and tortuous the sequences, to the laws of physics."

It is apparent from these passages that we are dealing here primarily with what we can call "methodological (or epistemic) reductionism" which concerns explanation of the phenomena of nature. Such reductionism, however, is usually based on "ontological (or metaphysical) reductionism" which seeks to define what is real, i.e. what the world is made of, and concludes (or posits) that the phenomena of nature are "*nothing but*" more or less complex agglomerations of the basic building blocks of physics, be they atoms, quarks, strings or whatever else the "final theory" will come up with.

Let me say that as a physicist I find this point of view quite familiar and in some sense implicit in the contemporary culture of science. Nevertheless, if I ask myself what precisely is meant by the words "explain", "may in some sense be reduced" or "ultimately", for example, in the above quotes, as well as the familiar concept of "nothing but" alluded to above, then it seems to me that this reductionism is very much like a credo, a faith concerning how the world is put together and how it can be understood. In that sense it is not so different from the statement "Christ died on the cross to save you and me, brother".

In my view the important issue is not so much whether you or I share this faith or a different one, but whether this grand reductionism is a fundamental paradigm of Science, as many would argue. I believe the answer is no, as explained below.

The next in the hierarchy of world views is what I will call *materialism*, which is also based on tangible observables, but which recognizes a difference between inanimate and living matter, in that it does not seek to reduce biology to physics. The distinction with reductionism is primarily methodological rather than ontological. It is here that the philosophical debate spills over into science, but the issues are immensely complex and resolution in the form of a clear Scientific consensus may be a long time in coming. Examples of the serious scientific debate can be found in Schrödinger (1945), Rosenberg (2006), Morris (2008), Kaufman (2008), Goldenfeld and Woese (2007) and I suspect that true scientific progress is what will illuminate the philosophical debate, rather than the other way around.

From an ontological point of view materialism still recognizes only matter as real, so that it considers thoughts, emotions or culture to be reducible to (nothing but) matter. In contrast to this point of view, we will designate as *naturalism* the view that gives to each category of phenomena its own ontological status and only requires methodological or logical *consistency*, i.e. explanations in one domain may not *contradict* the laws in another domain. Naturalism also strongly seeks to unify the different domains but not at the cost of reduction to one "fundamental" domain.

Needless to say these words are easier to write down than to explain in any illuminating way. The vast area of philosophy (and of science, if not Science) known as the mind-body problem is a testament to the richness of the issues, to which Nagel himself has contributed importantly (Nagel, 2004).

In our view naturalism as thus defined is the proper basis for Science (see Hohenberg, 2010). In practice the main consequence for Science is the rejection of *supernaturalism*, which allows principles or laws at one level to modify or contradict phenomena at another level and thereby explain them by a "higher authority". It should be said that according to the definition of Science given above or in Hohenberg (2010), supernaturalism could in principle have emerged as a basic paradigm of Science, since the ultimate criterion is what has gained the status of public knowledge, but in practice it has not done so in the past 400 years, so it is our view that naturalism is firmly part of the basic paradigm of Science which has been accepted either explicitly or implicitly.

Returning to the philosophical debate between Dawkins and Nagel, we see that Dawkins obviously embraces physicalist reductionism and Nagel objects, in our view quite correctly, that this is an unnecessarily dogmatic and restrictive point of view. Nagel himself does not adopt any specific approach in the classification listed above, appealing at one point to what we have termed (non-physicalist) materialism, as exemplified in the quote concerning teleological principles given at the beginning of this subsection. In Nagel (2008), on the other hand, he refers to the "spiritual hypothesis" or to the "… possibility that a nonphysical being should intervene in the natural order…", a point of view which we term supernaturalism, and he wonders why such hypotheses are not considered part of science. My answer is that empirically they are certainly not part of Science, at least in the past 400 years. To the question of whether they could form a legitimate subject for scientific inquiry, i.e. be part of (lower case) science, my answer is "Good luck trying to engage in a serious scientific debate on this basis. Try to send your work out into the scientific world and see if anyone picks up on it, a necessary precursor to emergence into Science". Four hundred years of intellectual history have taught us that this is a scientific dead end and I believe it is safe to bet that this will remain the case for the next 50 to 100 years at the very least.

V.     The Teaching of Evolution

Although there is much to say about the teaching of evolution in the schools and the National Academy of Sciences (2008), as well as many prominent scientists (see Ayala, 2007 ),  and public figures have weighed in, I shall primarily confine my discussion to a provocative article by Nagel (2008), which runs counter to the prevailing orthodoxy among educated persons. To quote his Introduction:

> "The political desire to defend science education against the threats of religious orthodoxy, understandable though it is, has resulted in a counter-orthodoxy, supported by bad arguments, and a tendency to overstate the legitimate scientific claims of evolutionary theory. Skeptics about the theory are seen as so dangerous, and so disreputably motivated, that they must be denied any shred of legitimate interest. Most importantly, the campaign of the scientific establishment to rule out intelligent design as beyond discussion because it is not science results in the avoidance of significant questions about the relation between evolutionary theory and religious belief, questions which must be faced in order to understand the theory and evaluate the scientific evidence for it."

The first distinction to be made is between the philosophical and scientific debates concerning evolution, on the one hand, and the proper way to teach evolution in high school science classes, on the other hand. The first question has two aspects: the general philosophical debate over reductionism, materialism and naturalism discussed above, and the spillover into evolutionary biology. These are immensely challenging issues that are not easily summarized, but my own belief with respect to the second question is that the next 20 to 50 years will show a significant shift away from physicalist reductionism in the scientific understanding of the interface between biology and physics. What it will be replaced with I do not know. Religion should be minimally affected by such a shift, except that one might hope that fundamentalist aberrations might be tempered. As has been said many times, intelligent design is primarily a political and religious movement, but to the extent that it is science (and ultimately anyone is free to engage in science), it is junk science which will disappear in due time. There is no chance that it will have any impact on Science.

Turning to the teaching of evolution, the essential aim should be to present it appropriately as a crowning achievement of Science without overselling the story. As such, the aspects to emphasize seem to me to be the following:

> • Evolution, like all great scientific theories is based on observation and logic, not on belief. Its main tenet is the relatedness of all forms of life, i.e. the Tree of Life, for which there is overwhelming evidence. Darwin's contribution is to explain the appearance of design by *natural selection*, the combination of genetic change and modification through heredity.

• As is true for all of Science the theory assumes *naturalism*, i.e. the rejection of supernatural explanations.

• Science does not address questions regarding the purpose of life or of events in the world, since the answers to such questions are likely to be diverse, i.e. different for different people and Science aims at universal answers.

• Most important, it should be stressed that evolutionary biology is a living science and that there are still many open unresolved questions, among which are the following:
  - What is the relationship between living and inanimate matter and between matter and thought, emotions, culture?
  - How should we understand and explore the origin of life?
  - Regarding natural selection, are the known and assumed biological mechanisms consistent with the time scale of evolution inferred from geological (i.e. physical) evidence?

From a tactical point of view I believe that it is important to emphasize first of all the fact of evolution as represented by the Tree of Life, ahead of the mechanism of evolution, natural selection. This is because the former is less controversial, easier to understand and on firmer empirical ground. That natural selection occurs cannot be doubted, but how it arose historically and how it functions quantitatively are still subjects of ongoing research. Basically, natural selection is not easy to understand and the theory is easily distorted and oversimplified. I am not suggesting that natural selection should be passed in silence but rather that it should be presented as secondary to the main story which is the observed fact of evolution, i.e. the Tree of Life.

It would also be useful to emphasize that Science is not based on grand reductionism or even on materialism, since that seems to be a flash point of sensitivity, though how to phrase this idea is a challenge. Finally, talk about purposelessness and the support that the theory of evolution gives to individuals' atheism should be considered self indulgences that serve no useful purpose and whose main effect is to inflame the debate.